\documentclass{elsart}

\usepackage{natbib}  
 
\begin{document}

% running author title:

\runauthor{Ulvestad}

% title, authors, acknowledgments: 

\begin{frontmatter} 
\title{Goals of the ARISE Space VLBI Mission\thanksref{sag}\thanksref{pub}}
\author[NRAO]{J.S. Ulvestad}

\address[NRAO]{National Radio Astronomy Observatory\thanksref{nrao}, 
Post Office Box O, 1003 Lopezville Road, Socorro, NM 87801-0387, USA}
 
\thanks[sag]{This paper is based
on the science ``white paper'' being developed
by the ARISE Science Advisory Group, and on the mission
study under way at NASA's Jet Propulsion Laboratory.}

\thanks[pub]{To be published in New Astronomy Reviews}

\thanks[nrao]{The National Radio Astronomy Observatory is operated
by Associated Universities, Inc., under a cooperative agreement
with the U.S. National Science Foundation.}

% an abstract (must be included!):

\begin{abstract} 
  
Supermassive black holes, with masses of
$10^6M_\odot$ to more than $10^9M_\odot$, are among the
most spectacular objects in the Universe, and
are laboratories for physics in extreme conditions.
The primary
goal of ARISE (Advanced Radio Interferometry between Space
and Earth) is to use the technique of Space VLBI to 
increase our understanding of black holes and their
environments, by imaging the havoc produced in the
near vicinity of the black holes by their enormous
gravitational fields.  The mission will be based on a 
25-meter space-borne radio telescope
operating at frequencies between 8 and 86~GHz,
roughly equivalent to an orbiting element of the
Very Long Baseline Array.  In an elliptical orbit with an
apogee height of 40,000--100,000~km, ARISE will provide
resolution of 15~microarcseconds or better, 5--10 times better
than that achievable on the ground.  At frequencies of 43 and 86~GHz,
the resolution of light weeks to light months in distant quasars
will complement the gamma-ray and X-ray observations of
high-energy photons, which come from the same regions
near the massive black holes.
At 22~GHz, ARISE will image the H$_2$O maser disks in active
galaxies more than 15~Mpc from Earth,
probing accretion physics and giving
accurate measurements of black-hole masses.  ARISE also will study
gravitational lenses at resolutions of tens of microarcseconds, yielding
important information on the dark-matter distribution and on
the possible existence of compact objects with masses of
$10^3M_\odot$ to $10^6M_\odot$.

\end{abstract} 

% enter no more than 6 relevant keywords (see the file keywords.txt 
% for a complete list) 


\end{frontmatter} 

\section{The ARISE Mission Concept}
\label{concept} 

Supermassive black holes (SMBHs) are thought to be responsible for the
astounding amount of energy released from the centers of
many galaxies.  The technique of Space VLBI \citep{U1} is
the only astronomical technique  
foreseen for the next 20~years that will have the capability of
imaging the region dominated by the gravitational
potential of the black hole, within light days to light months of
the active galactic nucleus.
ARISE (Advanced Radio Interferometry between Space
and Earth) is a mission currently under active study in the
U.S. that will orbit a 25-m telescope to work together
with ground telescopes worldwide in order to investigate
the spectacular astrophysics in the vicinity of SMBHs.

For ground-based VLBI operating at a frequency of 86~GHz on
the longest baselines possible on Earth ($\sim$10,000~km),
the best angular resolution is about 75~$\mu$as,
a factor of $\sim 500$ better than that achievable with the
Hubble Space Telescope.
ARISE, in an elliptical orbit with a maximum altitude
of 40,000-100,000~km, will
work together with sensitive ground radio telescopes
such as those in the Very Long Baseline Array (VLBA) and in the European
VLBI Network (EVN), and will produce radio images of active
galactic nuclei (AGNs) with angular
resolution of 7--15~$\mu$as
at the highest observing frequency of 86~GHz.
Table~\ref{mission} lists the basic mission parameters.

\begin{table}[hb!]
\begin{center}
\caption{ARISE Mission Parameters}
\label{mission}
\vspace{0.5cm}
\begin{tabular}{lc}
\hline
Parameter & Value \\
\hline
Apogee Height &40,000--100,000 km \\
Orbital Period&13--37 hr \\
Observing Bands&8, 22, 43, 86 GHz \\
Antenna Diameter&25 meters \\
Maximum Data Rate&4--8 Gbit sec$^{-1}$ \\
Polarization&Dual Circular or Linear \\
Maximum Baseline&4--9 Earth Diameters \\
\hline
\end{tabular}
\end{center}
\end{table}

Table~\ref{observ} lists the observation characteristics as
a function of frequency.  Detection thresholds for a baseline
to the Effelsberg 100-m telescope (EB) are given, assuming no
phase referencing and the maximum data rate.  Because of
angular momentum constraints, it is extremely unlikely that
the space radio telescope can switch sources rapidly enough to
do phase referencing, but this technique of calibrating the
atmosphere may be enabled just by having the ground telescopes
switch sources.  At 43 and 86 GHz, millimeter-wave telescopes
such as SEST and the MMA/LSA will be important anchors that
will significantly improve the fringe-detection threshold.

\begin{table}[ht!]
\begin{center}
\caption{Frequency-Dependent Observational Parameters for ARISE}
\label{observ}
\vspace{0.5cm}
\begin{tabular}{lcccc}
\hline
Parameter&8 GHz&22 GHz&43 GHz&86 GHz \\
\hline
Aperture Efficiency&0.50& 0.38& 0.24& 0.08 \\
System Temperature&12 K& 16 K& 24 K& 45 K \\
Resolution&$\leq 150$~$\mu$as&$\leq 60$~$\mu$as& $\leq 30$~$\mu$as&
$\leq 15$~$\mu$as \\
Detection Threshold to EB&0.7 mJy&1.9 mJy&9.4 mJy&150 mJy \\
\hline
\end{tabular}
\end{center}
\end{table}

\section{ARISE Science Goals}
\label{goals}

ARISE is a versatile, high-sensitivity instrument that will
employ the technique of Space VLBI 
to image the environment of a variety of compact objects 
such as supermassive
black holes (SMBHs).  It will resolve details 5--10 times smaller than
can be imaged using ground-based VLBI, and several
orders of magnitude smaller than instruments observing in other wavebands.
Table~\ref{scigoals} summarizes the primary science goals of ARISE;
a number of additional goals, such as imaging of young supernovae,
are omitted due to lack of space.
 
\begin{table}[ht!]
\begin{center}
\caption{Primary Science Goals of ARISE}
\label{scigoals}
\vspace{0.5cm}
\begin{tabular}{l}
\hline
\underbar{Supermassive Black Holes and Radio Jets} \\
\ \ \ \ \ AGN Fueling \\
\ \ \ \ \ Relativistic Jet Production \\
\ \ \ \ \ Generation of High-Energy Photons \\
\underbar{Accretion Disks and H$_2$O Megamasers} \\
\ \ \ \ \ Masses of Supermassive Black Holes \\
\ \ \ \ \ Nature of Megamaser Disks \\
\ \ \ \ \ Accretion Processes \\
\ \ \ \ \ Geometric Distance Measurements \\
\underbar{Cosmology} \\
\ \ \ \ \ Gravitational Lens Studies \\
\ \ \ \ \ High-Redshift Radio Sources \\
\hline
\end{tabular}
\end{center}
\end{table}

The most important goals of ARISE focus on studies of SMBHs
and their environments in active galactic nuclei,
the most energetic power plants in the Universe. The popular treatment by
\citet{B1} discusses observed properties of AGNs over a variety of
wavebands that are attributable to SMBHs.
The current paradigm for an AGN includes, at
its center, a SMBH that provides the power for the AGN.
Surrounding the black hole is an accretion disk,
that is roughly co-planar (except for disk warps) with a much more
extensive ``torus'' of material that may extend for hundreds
of parsecs.  As material in the disk drifts toward the
central black hole, energy is extracted from that material
by the spinning black hole.
A magnetized radio jet of highly relativistic particles is
accelerated near the SMBH, and flows outward near the speed
of light along the symmetry axis of the accretion disk.
Flickering gamma-ray emission reveals the
creation of large quantities of high-energy particles in the
inner light months of the radio jet.
 
With ARISE, two critical
classes of observations can be made.  First, imaging
of the inner light months of active galaxies in their
continuum radio emission reveals the birthplace
of the relativistic jets, the generation of shocks
near that birthplace, and the key physical parameters
in the regions of gamma-ray production.  Second,
imaging of molecular line (H$_2$O maser) emission from the
inner light months of the accretion disks
in AGN directly samples the dynamics of material in the
vicinity of the SMBH.  Such studies lead to direct measurement
of SMBH masses and of the physical characteristics of the
accretion process \citep{M1}.
VLBI in general, and ARISE in
particular, provide important information, and actual images,
that can be supplied by no other technique in modern
astrophysics.  The ARISE resolution at 86~GHz will correspond
to $\leq 0.1$~pc for a blazar at $z=0.5$, enabling resolution
on a scale similar to that of the gamma-ray emission.
In an H$_2$O megamaser galaxy at 50~Mpc distance, the
22-GHz resolution will be $\leq 0.05$~pc, enabling
imaging of the vertical and velocity structures in the
disk.  Observations on such important physical scales in
these objects are not possible with VLBI baselines whose
length is limited by the size of the Earth.

Beyond the studies of SMBHs and their environment, ARISE
can use AGNs for a variety of important cosmological studies.
In particular, ARISE will permit investigation of radio
sources with an otherwise
unreachable combination of sensitivity and angular resolution,
which is crucial for conclusive cosmological tests measuring
the dependence of angular size and separation on redshift.
Of special interest are the novel investigations that can
be made using gravitational lenses \citep{K1}. 
ARISE imaging of lensed AGNs will 
improve the modeling of the mass distribution,
currently the largest uncertainty in the determination
of the Hubble Constant by this direct method.
A gravitational lens also acts as a
``cosmic telescope'' in magnifying the background source by
a factor of 10 or more, effectively increasing the
angular resolution of ARISE to near 1~$\mu$as, which will
provide resolution of light days even for the most distant AGNs.
Finally, the sensitivity of ARISE
to structures on the scale of tens to hundreds of microarcseconds
will enable detection of compact lenses having masses of
$10^3$--$10^6M_\odot$; such objects
are among the leading candidates for the ``missing'' baryonic dark matter.

\section{European Contributions to ARISE}

ARISE is currently part of the long-term roadmap in NASA's
Structure and Evolution of the Universe theme; if ARISE is
funded in its current incarnation, it will likely
be as a U.S.-led mission.  However, ARISE also has many of
the elements of a ``descendant'' of two other Space VLBI 
concepts, QUASAT and the International VLBI Satellite, which
were proposed to the European Space Agency (ESA), but ultimately
were not funded.  Thus, concepts developed in Europe already
have played a key role in ARISE, and several members of the
ARISE Science Advisory Group are based at European institutions.
The newly formed Joint Institute for VLBI in Europe (JIVE)
provides an excellent vehicle for
the participation of European ground facilities in ARISE;
the European development of millimeter-wave telescopes will
be especially useful at the higher frequencies.
The capabilities of ESA, including equipment that will be
flown aboard Planck/FIRST, also could provide important 
contributions to the space element of ARISE.  Table~\ref{europe}
lists some areas in which European participation could contribute
significantly to ARISE.

\begin{table}[hb]
\begin{center}
\caption{Possible European Participation in ARISE}
\label{europe}
\vspace{0.5cm}
\begin{tabular}{l}
\hline
Science planning \\
Ground radio telescopes \\
Recording systems \\
JIVE correlator \\
Tracking station \\
On-board VLBI equipment \\
Space cryo-cooler or amplifiers \\
\hline
\end{tabular}
\end{center}
\end{table}

\section{ARISE Timeliness}

The VLBI Space Observatory Programme (VSOP), is the first dedicated
Space VLBI mission, in operation since early 1997 \citep{H1}.  VSOP,
under the leadership of the Institute for Space and Astronautical
Science in Japan, has 
demonstrated the capability for routine Space VLBI imaging by
observing strong sources at 1.6 and 5~GHz.
A much more sensitive mission using this
technique will be timely, because it provides an imaging capability
in the compact regions that will be investigated by several
upcoming high-energy satellites such as GLAST and ASTRO-E.
ARISE will take advantage of space technologies currently under 
active development.  The most crucial technology is that connected
with the deployable 25-m reflector that must work at frequencies
as high as 43 and 86~GHz.  The current baseline selection for
ARISE is an inflatable antenna, under development for
several other applications in communications and remote
sensing.  The other ``new'' technologies are those aimed at
achieving a very high sensitivity, and should be well in hand
by the potential ARISE launch date of 2008.  These include
low-noise amplifiers (developed for MAP and Planck/FIRST) and
cryogenic cooling to an ambient temperature of 20~K 
(tested aboard the Space Shuttle and required for Planck/FIRST).

Required ground systems include a number of sensitive
ground telescopes.  The EVN and the VLBA already
provide a suite of ground telescopes as well as the entire
operational infrastructure necessary for a VLBI mission.
Completion of the Green Bank Telescope, the MMA/LSA, and the 
VLA upgrade will provide major new capabilities at the
highest ARISE observing frequencies.  Finally 
multi-gigabit per second data-recording and correlation
capability will be required, and is under active development
by several groups, notably in the Mark~4 and S3/S4 systems.
Within the U.S., ARISE provides a unique opportunity for
cooperation between space assets funded by NASA and an
extensive ground infrastructure already developed under
funding by the National Science Foundation.  Thus, ARISE is
a timely mission because it can take advantage of the large
investments already made in ground facilities in both the
U.S. and Europe.

\end{document}